\documentclass[10pt, aps,prl,twocolumn,amsmath,amssymb,superscriptaddress]{revtex4-2}

\usepackage{graphicx} % Required for inserting images

\newcommand{\LL}[1]{\textcolor{black}{#1}} 
\newcommand{\AM}[1]{\textcolor{black}{#1}} 

\usepackage[utf8]{inputenc}
\usepackage{lipsum}
\usepackage{svg}
\usepackage{graphicx}
\usepackage{braket}
\usepackage{tabularx}
\usepackage{textcomp, gensymb}
\usepackage{color}
\usepackage{comment}
\usepackage{lineno}

\begin{document}
	
	\title{\textbf{Spin-to-polarization mapping with a coherent quantum dot-cavity receiver}}
	\author{A. Medeiros}
	\affiliation{Université Paris-Saclay, CNRS, Centre de Nanosciences et de Nanotechnologies, 91120, Palaiseau, France}
	\author{V. Vinel}
	\affiliation{Université Paris-Saclay, CNRS, Centre de Nanosciences et de Nanotechnologies, 91120, Palaiseau, France}
	\author{E. Rambeau}
	\affiliation{Université Paris-Saclay, CNRS, Centre de Nanosciences et de Nanotechnologies, 91120, Palaiseau, France}
	\affiliation{Universit\'e Paris Cit\'e, CNRS, Centre de Nanosciences et de Nanotechnologies, F-91120 Palaiseau, France} 
	\author{P. Steindl}
	\affiliation{Université Paris-Saclay, CNRS, Centre de Nanosciences et de Nanotechnologies, 91120, Palaiseau, France}
	\author{E. Mehdi}
	\affiliation{Université Paris-Saclay, CNRS, Centre de Nanosciences et de Nanotechnologies, 91120, Palaiseau, France}
	\affiliation{Universit\'e Paris Cit\'e, CNRS, Centre de Nanosciences et de Nanotechnologies, F-91120 Palaiseau, France}
	\author{M. Gundín}
	\affiliation{Université Paris-Saclay, CNRS, Centre de Nanosciences et de Nanotechnologies, 91120, Palaiseau, France}
	\affiliation{ICFO-Institut de Ciencies Fotoniques, The Barcelona Institute of Science and
		Technology, Castelldefels (Barcelona), 08860, Spain.}
	\author{C. Millet}
	\affiliation{Université Paris-Saclay, CNRS, Centre de Nanosciences et de Nanotechnologies, 91120, Palaiseau, France}
	\author{P. Stepanov}
	\affiliation{Quandela, 7 Rue Léonard de Vinci, 91300 Massy, France}
	\author{N. Somaschi}
	\affiliation{Université Paris-Saclay, CNRS, Centre de Nanosciences et de Nanotechnologies, 91120, Palaiseau, France}
	\affiliation{Quandela, 7 Rue Léonard de Vinci, 91300 Massy, France}
	\author{A. Lemaître}
	\affiliation{Université Paris-Saclay, CNRS, Centre de Nanosciences et de Nanotechnologies, 91120, Palaiseau, France}
	\author{I. Sagnes}
	\affiliation{Université Paris-Saclay, CNRS, Centre de Nanosciences et de Nanotechnologies, 91120, Palaiseau, France}
	\author{O. Krebs}
	\affiliation{Université Paris-Saclay, CNRS, Centre de Nanosciences et de Nanotechnologies, 91120, Palaiseau, France}
	\author{P. Senellart}
	\affiliation{Université Paris-Saclay, CNRS, Centre de Nanosciences et de Nanotechnologies, 91120, Palaiseau, France}
	\author{D. A. Fioretto}
	\affiliation{Université Paris-Saclay, CNRS, Centre de Nanosciences et de Nanotechnologies, 91120, Palaiseau, France}
	\author{L. Lanco}
	\affiliation{Université Paris-Saclay, CNRS, Centre de Nanosciences et de Nanotechnologies, 91120, Palaiseau, France}
	\affiliation{Universit\'e Paris Cit\'e, CNRS, Centre de Nanosciences et de Nanotechnologies, F-91120 Palaiseau, France} 
	%\affiliation{Institut Universitaire de France (IUF)}
	
	\begin{abstract}
		Coherent light-matter interfaces controllably modifying the state of a photon upon interaction with a stationary qubit are a key resource for implementing deterministic \LL{entangling} gates for optical quantum technologies. This requires a one-to-one mapping between the state of the scattered photon and that of the embedded qubit. Here, we present an experimental signature of such a bijection by leveraging the spin-induced Kerr rotation present in a low-noise charged quantum dot-micropillar cavity device. Through time-resolved polarization measurements, we project the electron spin to one of its eigenstates with $95\pm2\%$ fidelity with a single reflected photon detection, and follow the subsequent spin relaxation through the detection of a second reflected photon. We demonstrate that, after a transient regime governed by the trion radiative lifetime, two orthogonal polarization states can be produced, each associated to a given spin eigenstate. While the current results are limited by a timescale competition between electron spin relaxation and trion radiative lifetime, they could be improved using hole spins displaying increased relaxation times. Our work paves the way towards deterministic logic gates exploiting this one-to-one mapping between a spin and the polarization of a scattered photon.
	\end{abstract}

	\maketitle
	Efficient light-matter interfaces are a key resource in the development of scalable photonic quantum technologies \cite{Reiserer2015, Borregaard2019}. \LL{This includes the development of} coherent receivers \LL{which} can be used to manipulate incoming photons, providing atom-induced rotations of the scattered photonic qubit state \cite{Duan2004}. Different platforms such as atoms \cite{Tiecke2014,Volz2014,Bechler2018,Daiss2021,Stolz2022}, vacancy centers \cite{Bhaskar2020,Pasini2024,Herrmann2024}, rare-earth ions \cite{Zhong2015} and semiconductor quantum dots (QDs) \cite{DeSantis2017,LeJeannic2022,Mehdi2024}\LL{, coupled to cavities or waveguides,} have been explored as such.
	
	\LL{Coherent} receivers have the potential to implement deterministic photon-photon gates \cite{Chang2014, Yang2022, Hu2008_Entangler}, which can be used both for gate-based quantum computing \cite{Wei2013} and for the deterministic generation of entangled resource states in measurement-based quantum computing \cite{Hu2008_Entangler, Pichler2017}. However, this requires engineering a one-to-one bijective mapping between orthogonal states of the scattered photonic qubit and orthogonal states of the atomic qubit \cite{Duan2004}. Contrary to emitters \LL{\cite{Thomas2022, Thomas2024, Meng2024, Coste2023_Cluster}}, such a mapping does not directly arise from optical selection rules but has to be engineered by harnessing the interference between the input field and the \AM{scattered light by the atom}.%atomic resonance fluorescence.
	
	In this context, solid-state spins have emerged as promising candidates for the development of scalable devices that can be coupled to optical cavities or \LL{nanophotonic waveguides} \cite{Islam2026}. Coherent spin-photon receivers, based on color centers or semiconductor QDs, have already been used to demonstrate quantum memories \cite{Nguyen2019, Stas2022}, remote spin-photon entanglement \cite{Bhaskar2020, Knaut2024, Chan2023} and spin-photon gates \cite{Sun2016, Sun2018}.
	
	In particular, semiconductor QDs embedded in micropillar cavities hold great promise \cite{hu2008, Bonato2010} due to their almost Fourier-limited linewidths \cite{Wang2016, DeSantis2017}, optical selection rules in polarization \cite{Coste2023} and giant spin-dependent Kerr rotations \cite{Mehdi2024}. \LL{The limited birefringence of pillar cavities allows directly exploiting polarization qubits, providing straightforward 1-qubit gates and measurements,
		as well as simple protocols for the implementation of
		various spin-photon and multi-photon gates \AM{\cite{Leuenberger2006, Bonato2010, Hu2008_Entangler, Lindner2009}}}. In such devices, the polarization of an incoming photon is modulated upon reflection due to the interference between the directly reflected light and the Rayleigh-scattered spin-dependent emission. This interaction can be tuned to engineer a bijective mapping whereby orthogonal spin states lead to orthogonal polarization states upon photon reflection \cite{Arnold2015}. Experimentally, the giant Kerr effect has been used to probe the relaxation dynamics of a single hole spin \cite{Gundin2025}\LL{, as well as the relaxation and decoherence dynamics of an electron spin in a transverse magnetic field \cite{Medeiros2026}. However}, a direct experimental signature of a one-to-one spin-to-polarization mapping has yet to be demonstrated.
	
	In this paper, we \AM{evidence} such a spin-to-polarization mapping using \AM{cavity-}enhanced Kerr rotations in a low-noise QD-cavity receiver embedding a single electron spin. We first implement a measurement-based heralding of the single spin state upon the detection of a reflected photon in a given polarization state. We demonstrate that spin projection can be quantified by performing time-resolved polarization measurements, as the spin relaxation dynamics is transferred onto the polarization state of a second photon. We use this scheme to show experimental signatures that low-noise QD-cavity devices can indeed provide opposite spin-dependent polarization states in the Poincaré sphere. The time-dependent conditional tomography \AM{measurement} also allows exploring the progressive build-up of the coherent reflectivity response, limiting the purity of the observed conditional polarization states when performing experiments with an electron spin with nanosecond relaxation time. Our results are supported by numerical simulations and an analytical semiclassical description.\\
	
	\noindent \textbf{\Large{Results}}
	
	\noindent\textbf{Principle of the experiments}
	
	The device under study is a negatively-charged InGaAs QD emitting at 925 nm, embedded in an AlGaAs micropillar cavity \AM{(see Methods for details on device fabrication)}. The QD energy levels are displayed, alongside a schematics of the device, in Fig.~\ref{fig:1_principle_experiment}a. The two electron spin states $\ket{\uparrow}_z$ and $\ket{\downarrow}_z$ form the ground states\LL{, with $z$ the growth axis}; the excited states are the two trion states $\ket{\Uparrow\uparrow\downarrow}_z$ and $\ket{\Downarrow\downarrow\uparrow}_z$, consisting of a pair of electrons in a singlet configuration and a single hole (we drop the $z$ sublabel from now on for clarity purposes). The states are coupled to light with circularly polarized optical selection rules \cite{Warburton2013}. \AM{The micropillar cavity, sketched in Fig.~\ref{fig:1_principle_experiment}b}, has two slightly birefringent orthogonally polarized modes $H$ and $V$, \AM{with quality factors $Q_H=7400$ and $Q_V=6900$, respectively}\LL{. The QD is coupled to both modes, though it is closer to resonance with the H-polarized one (see insert of Fig.~\ref{fig:1_principle_experiment}b)}. Both QD transitions are addressed with an $H$-polarized \LL{continuous-wave (CW) }laser in the low-power regime at frequency $\omega$. The spin-dependent polarization of a reflected photon will result from the interference between the light directly reflected from the top mirror and the QD emission \cite{Arnold2015, Mehdi2024}.
	
	\AM{We first introduce a model valid in the low-power limit, when the population of the excited trion states can be neglected~\cite{Steindl2023}, and in the absence of decoherence both for the optical transitions and for the spin states, considered stable.} We call $\ket{\Psi_\uparrow}$ (resp.  $\ket{\Psi_\downarrow}$) the pure polarization state obtained for the reflected light if the spin is $\ket{\uparrow}$ (resp. $\ket{\downarrow}$), with:
	\begin{equation}
		\ket{\Psi_{\uparrow/\downarrow}}=\frac{r_{H\rightarrow H}\ket H\pm r_{H\rightarrow V}\ket V}{\sqrt{|r_{H\rightarrow H}|^2+|r_{H\rightarrow V}|^2}},
		\label{eq:Ideal_Polarization_States}
	\end{equation}
	with a + (resp. -) sign if the spin is $\ket\uparrow$ (resp. $\ket\downarrow$). Here, $r_{H\rightarrow H}$ (resp. $r_{H\rightarrow V}$) is the reflection amplitude associated to detecting a photon in polarization $H$ (resp.~$V$), under an $H$-polarized input excitation. These amplitudes depend on the device parameters \AM{(see Methods)} and on the detuning $\Delta\omega$ between the input field $\omega$ and the QD resonance frequency $\omega_{\rm QD}$ \cite{Arnold2015}. In such a model, the pure polarization states $\ket{\Psi_\uparrow}$ and $\ket{\Psi_\downarrow}$ can be coherently rotated by experimentally tuning $\Delta\omega$ \cite{Arnold2015, Mehdi2024}. The overlap between these two states is described by the scalar product $|\braket{\Psi_\uparrow|\Psi_\downarrow}|^2$:
	\begin{equation}
		|\braket{\Psi_\uparrow|\Psi_\downarrow}|^2=\left(\frac{|r_{H\rightarrow H}|^2-|r_{H\rightarrow V}|^2}{|r_{H\rightarrow H}|^2+|r_{H\rightarrow V}|^2}\right)^2=s_{HV}^2,
		\label{eq:Ideal_Scalar_product}
	\end{equation}
	where $s_{HV}$ is the Stokes parameter describing the intensity contrast in the $H-V$ axis of the Poincaré sphere~\cite{Hilaire2018}. Fig.~\ref{fig:1_principle_experiment}c shows the analytically predicted scalar product $\left|\braket{\Psi_\uparrow|\Psi_\downarrow}\right|^2$ as a function of the detuning $\Delta\omega$ for the parameters \AM{characterizing this device} \LL{(see Methods)}. For large detunings% ($|\Delta\omega|\gg0$)
	, the light is \AM{reflected in the $H$ polarization, since no light is scattered by the QD spin}, hence $|r_{H\rightarrow H}|\sim1$ and $|r_{H\rightarrow V}|\sim 0$, leading to $|\braket{\Psi_\uparrow|\Psi_\downarrow}|^2\sim1$. As the laser is tuned in resonance with the QD% ($|\Delta\omega|\sim0$)
	, the polarization is rotated in a spin-dependent way, with $V$-polarized light now being \AM{scattered}, leading to an increase of $|r_{H\rightarrow V}|^2$. In particular, there are two detunings for which $|r_{H\rightarrow H}|=|r_{H\rightarrow V}|$ ($\braket{\Psi_\uparrow|\Psi_\downarrow}=0$, i.e. $s_{HV}=0$). This is the condition for an optimal interface, with orthogonal spin-dependent polarization states. In this specific configuration where $s_{\rm HV}=0$, Eq.~(\ref{eq:Ideal_Polarization_States}) simplifies to:
	\begin{figure}[t!]
		\centering
		\includegraphics[scale=1]{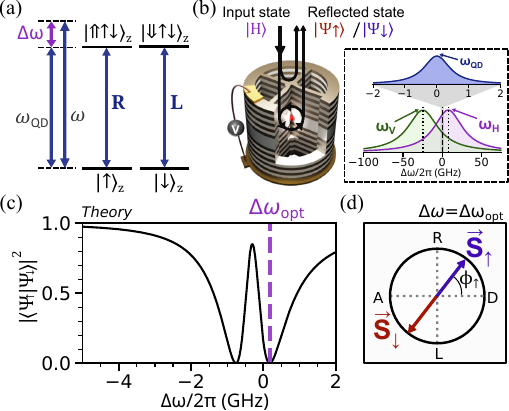}
		\caption{(a) Energy levels and optical selection rules of a negatively charged QD. The device is optically addressed with an input laser beam of frequency $\omega$, detuned by $\Delta\omega$ from the QD resonant frequency $\omega_{\rm QD}$. (b) An $H$-polarized input state is sent to the device. Upon reflection, its polarization is rotated in a spin-dependent way, leading to pure spin-dependent polarization states $\ket{\Psi_\uparrow}$ or $\ket{\Psi_\downarrow}$ if decoherence is neglected. Inset: Spectral configuration of the QD-cavity system. (c) Theoretically predicted scalar product $|\braket{\Psi_\uparrow|\Psi_\downarrow}|^2$ for the device under study, in the absence of decoherence, as a function of $\Delta\omega$. The dashed purple line indicates an optimal detuning $\Delta\omega_{\rm opt}$ where $\braket{\Psi_\uparrow|\Psi_\downarrow}=0$. (d) Stokes vectors $\vec S_\uparrow$ and $\vec S_\downarrow$ at $\Delta\omega_{\rm opt}$. The vectors are shown to be antiparallel, i.e. corresponding to orthogonal polarization states with an axis defined by the angle $\phi_\uparrow$.}
		\label{fig:1_principle_experiment}
		\centering
	\end{figure}
	\begin{equation}
		\ket{\Psi_{\uparrow/\downarrow}}=\frac{1}{\sqrt2}\left(\ket H \pm e^{i\phi_\uparrow}\ket V\right)
		\label{eq:ideal_psiup_psidown}
	\end{equation}
	
	where $\phi_\uparrow$ is the relative phase between the reflection amplitudes $r_{H\rightarrow H}$ and $r_{H\rightarrow V}$. In the Stokes vector representation, the \LL{polarization states $\ket{\Psi_\uparrow}$ and $\ket{\Psi_\downarrow}$ are associated to the vectors \mbox{$\vec S_\uparrow$} and $\vec S_\downarrow$ with $\vec S_\uparrow=-\vec S_\downarrow$ and \mbox{$\|\vec S_\uparrow\|=\|\vec S_\downarrow\| = 1$}.} Fig.~\ref{fig:1_principle_experiment}d shows these ideal $\vec S_\uparrow$ and $\vec S_\downarrow$ vectors, as predicted for the optimal detuning $\Delta\omega=\Delta\omega_{\rm opt}$ (vertical dashed line, in Fig.~1c), in the  \LL{plane containing the $D$, $A$, $R$, $L$ states} of the Poincaré sphere\AM{, where $s_{HV}=0$ is satisfied.}
	
	This bijection between the $\ket{\Psi_\uparrow}$ and $\ket{\Psi_\downarrow}$ polarization states \AM{and the $\ket\uparrow$ and $\ket\downarrow$ spin states} can be leveraged to herald the resident QD spin orientation, starting from a mixed state. In the following, we consider the case where \LL{a} reflected photon has been measured in a state \mbox{$\ket\phi\sim\ket H+e^{i\phi}\ket V$}, i.e. a pure polarization state in the DARL plane of the Poincaré sphere defined by an angle $\phi$ with respect to the $D$-polarization (see Fig.~\ref{fig:1_principle_experiment}d). Such a measurement leads to a back-action onto the spin state, with a post-measurement spin density matrix described by $\rho_{|\phi}$, conditioned to the detection of a reflected photon in the polarization state $\ket\phi$. \LL{Here, as well as in the following, the subscript $|\phi$ is used to describe conditional quantities, conditioned to the detection of a $\phi$-polarized photon}. In the optimal interface configuration, where $\ket{\Psi_\uparrow}$ and $\ket{\Psi_\downarrow}$ are described by Eq.~(\ref{eq:ideal_psiup_psidown}), the conditional spin density matrix $\rho_{|\phi}%^{\rm (s)}
	$ is:
	\begin{equation}
		\rho_{|\phi}%^{\rm (s)}
		=P_{\uparrow|\phi}(\tau)\ket\uparrow\bra\uparrow+P_{\downarrow|\phi}(\tau)\ket\downarrow\bra\downarrow,
		\label{eq:post_measurement_state}
	\end{equation}
	with $P_{\uparrow|\phi}(0)$ (resp. $P_{\downarrow|\phi}(0)=1-P_{\uparrow|\phi}(0)$) the conditional probability of finding the spin in the $\ket\uparrow$ (resp. $\ket\downarrow$) state after the measurement. In the case of an ideal interface with \LL{$\ket{\Psi_\uparrow}$ and $\ket{\Psi_\downarrow}$ described by Eq.~(\ref{eq:ideal_psiup_psidown})}%  $\braket{\Psi_\uparrow|\Psi_\downarrow}=0$
	, these probabilities are  \mbox{$P_{\uparrow|\phi}(0)=\frac{1}{2}\left(1+\cos(\phi-\phi_\uparrow)\right)$} and \mbox{$P_{\downarrow|\phi}(0)=\frac{1}{2}\left(1-\cos(\phi-\phi_\uparrow)\right)$} \AM{(see Supplemental Materials, hereafter denoted as \cite{Suppl})}. Based on these probabilities, the conditional Bloch component along the $z$-spin eigenaxis, after the photon is measured in the polarization state $\ket\phi$ at $\tau=0$, is:
	\begin{equation}
		\braket{\hat\sigma_z}_{|\phi}(0)=P_{\uparrow|\phi}(0)-P_{\downarrow|\phi}(0)=\cos(\phi-\phi_\uparrow).
		\label{eq:spin_predicted}
	\end{equation}
	\AM{As such, if $\phi=\phi_\uparrow$ (i.e. $\ket\phi=\ket{\Psi_\uparrow}$) the spin is maximally projected onto the $\ket\uparrow$ state right after the measurement. If a second photon is reflected at a time delay~$\tau$ after the first, the measured polarization will be $\phi=\phi_\uparrow$. In the presence of spin relaxation, however, the polarization will change following the spin decay towards the mixed state. The same reasoning applies if $\phi=\phi_\downarrow=\phi_\uparrow + \pi$ (i.e. $\ket\phi=\ket{\Psi_\downarrow}$), where the spin is projected in the $\ket\downarrow$ state.}
	
	\LL{Experimentally, one }\AM{can probe} both spin projection and the subsequent relaxation by measuring a second reflected photon \LL{along the $\phi\bar\phi$ axis of the Poincaré sphere, joining the states $\ket\phi$ and \mbox{$\ket{\bar\phi}\sim\ket H-e^{i\phi}\ket V$}, }orthogonal to $\ket\phi$. \LL{The conditional Stokes parameter measuring the second photon's polarization along this $\phi\bar\phi$ axis, conditioned to a first photon measured in state $\ket\phi$, is denoted as $s_{\phi\bar\phi|\phi}(\tau)$. This conditional Stokes parameter can be measured as a function of the delay $\tau$ between the two photon detection events} as:
	\begin{equation}
		s_{\phi\bar\phi|\phi}(\tau)=\frac{P_{\phi|\phi}(\tau)-P_{\bar\phi|\phi}(\tau)}{P_{\phi|\phi}(\tau)+P_{\bar\phi|\phi}(\tau)}.
	\end{equation}
	Here, $P_{\phi|\phi}(\tau)$ (resp. $P_{\bar\phi|\phi}(\tau)$) is the conditional probability to measure a second photon in polarization state $\ket\phi$ (resp. $\ket{\bar\phi}$) at a time delay $\tau$ after measuring the first one in $\ket\phi$. %In our simplified model neglecting decoherence, and 
	In the specific case of an ideal spin-photon interface $\braket{\Psi_\uparrow|\Psi_\downarrow}=0$ with an initially unpolarized spin, the conditional photon polarization and the conditional spin polarization are related through \cite{Suppl}:
	\begin{equation}
		s_{\phi\bar\phi|\phi}(\tau)=\cos^2(\phi-\phi_\uparrow)=\braket{\hat\sigma_z}_{|\phi}^2,
		\label{eq:spin_photon_correspondence}
	\end{equation}
	\AM{If $\phi=\phi_\uparrow$ (resp. $\phi=\phi_\downarrow=\phi_\uparrow+\pi$), the spin is projected in the $\ket\uparrow$ state, i.e. $\braket{\hat\sigma_z}_{|\phi}=1$ (resp. $\ket\downarrow$, i.e. $\braket{\hat\sigma_z}_{|\phi}=-1$). In the absence of spin relaxation, a second photon will be scattered with the same polarization state $\ket\phi=\ket{\Psi_\uparrow}$ (resp. $\ket{\Psi_\downarrow}$), leading to $s_{\phi\bar\phi|\phi}=1$. If $\phi=\phi_\uparrow\pm\pi/2$, the spin remains unpolarized after photon detection, i.e. $\braket{\hat\sigma_z}_{|\phi}=0$, and thus  \mbox{$s_{\phi\bar\phi|\phi}=0$}.}\\
	
	\noindent\textbf{Experimental setup and device characterization}
	
	In the following, we describe how we experimentally access the conditional probabilities $P_{\phi|\phi}(\tau)$ and $P_{\bar\phi|\phi}$ through two-photon correlations where the two detection events are either in the same polarization state (auto-correlations) or in orthogonal states (cross-correlations).
	
\begin{figure}[t!]
		\includegraphics[scale=1]{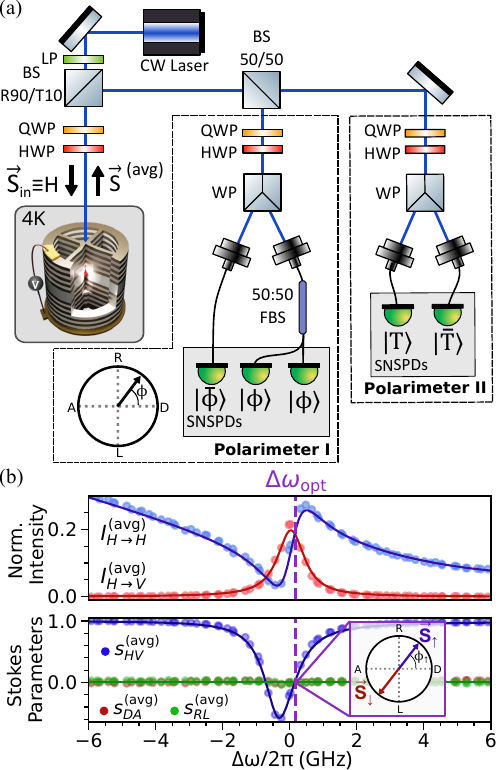}
		\caption{(a) Optical setup used to project the spin state. Two polarization analyzers measure the reflected light, in state $\vec S^{\rm (avg)}$. Polarimeter I measures light in polarization state $\ket \phi$ or $\ket{\bar\phi}$, with $\braket{\phi|\bar\phi}=0$, where $\phi=0$ corresponds to the $D$ state. A fiber BS splits the light in order to measure the same polarization state $\ket\phi$ in two different detectors. Polarimeter II performs the polarization tomography of the reflected light, using three mutually-orthogonal polarization bases $T\bar T=HV,~DA,~RL$. BS: Beamsplitter, FBS: Fiber beamsplitter, QWP: Quarter Wave Plate, HWP: Half Wave Plate, LP: Linear Polarizer, WP: Wollaston Prism, SNSPD: Super-conducting Nanowire Single Photon Detectors.
			(b) \AM{Polarization-dependent reflectivity measurements in the stationary regime as a function of $\Delta\omega$.} Top: The co-polarized average intensity $I_{H\rightarrow H}^{\rm (avg)}$ (blue) and cross-polarized average intensity $I_{H\rightarrow V}^{\rm (avg)}$ (red) are shown as a function of the laser-QD detuning.
			Bottom: Polarization-tomography of the stationary regime. The three average Stokes parameters $s_{HV}^{\rm (avg)}$ (blue), $s_{DA}^{\rm (avg)}$ (red) and $s_{RL}^{\rm (avg)}$  (green) are shown as a function of the laser-QD detuning. The vertical red line denotes the $\Delta\omega_{\rm opt}$ condition, corresponding to $s_{HV}^{\rm (avg)}=0$. The experimental data (scatter plot) are compared with numerical simulations (solid lines). }
		\label{fig:2_exp_setup_characterization}
		\centering
	\end{figure}

	The optical setup used to measure conditional probabilities and conditional Stokes parameters is schematized in Fig.~\ref{fig:2_exp_setup_characterization}a. The polarization of the input light, of varying frequency $\omega$, is fixed with a linear polarizer (LP) and modified with a set of half- and quarter- waveplates (HWP/QWP) to address only the cavity mode $H$ (input Stokes vector $\vec S_{\rm in}\equiv H$). The laser is focused into the micropillar cavity within a cryostat operating at 4K \cite{Mehdi2024}. The reflected light % (average output Stokes vector $\vec S^{\rm (avg)}$ \textcolor{purple}{[Eliott mentioned this comes out of nowhere]})
	is separated by a R90/T10 non-polarizing beam splitter (BS) and spatially filtered in a single mode fiber (not shown). A 50/50 free space BS divides the signal into two polarimeters. Polarimeter I measures light in the polarization state $\ket\phi\sim\ket H+e^{i\phi}\ket V$ and its orthogonal state $\ket{\bar\phi}\sim\ket H-e^{i\phi}\ket V$, where $\phi=0$ corresponds to the $D$ state (see Left inset). For this work, 75 measurement bases were calibrated with $\phi\in(-15\degree,150\degree)$. A 50/50 fiber BS is placed at the output of one of the collimators to measure \AM{auto-correlations with} the same polarization state $\ket\phi$ in two different detectors. Polarimeter II performs the polarization tomography of the reflected light by measuring it in three orthogonal bases ($HV,~DA,~RL$), to reconstruct the full reflected polarization state. All outputs for all polarimeters are connected to superconducting nanowire single photon detectors (SNSPD) and correlated via a time-correlator.
	
	\AM{As the spin is initially in a mixed state, the optical response results from the average between the $\vec S_\uparrow$ and $\vec S_\downarrow$ vectors, which we access by studying the steady-state optical response of the device, as a function of the laser-QD detuning $\Delta\omega$.} Fig.~\ref{fig:2_exp_setup_characterization}b (Top) shows polarization-resolved intensities in the $H$ and $V$ bases as a function of the detuning. The peak in the \AM{averaged} cross-polarized intensity $I_{H\rightarrow V}^{\rm (avg)}$ arises from the cross-polarized QD resonance fluorescence, while the signal in the \AM{averaged} co-polarized intensity $I_{H\rightarrow H}^{\rm (avg)}$ results from the interference between the input field and the co-polarized QD resonance fluorescence~\cite{Mehdi2024}. In this and all subsequent figures, we compare experimental data to a numerical model \cite{Suppl} taking into account optical pure dephasing, electron hyperfine interaction and charge cotunelling, \LL{describing all the observed features with a single set of parameters given in the Methods section}. From the parameters, we extract cooperativities $C_H=1.4\pm0.2$ and $C_V=1.3\pm0.2$. With $C_H,C_V>1$, our device should allow reaching the optimal spin-photon interface condition, i.e. the orthogonality of the spin-dependent polarization states \cite{Arnold2015}.
	
	Fig.~\ref{fig:2_exp_setup_characterization}b (Bottom) shows the polarization state of the average Stokes vector $\vec S^{\rm (avg)}$ through the three Stokes parameters $s_{HV}^{\rm (avg)}$, $s_{DA}^{\rm (avg)}$ and $s_{RL}^{\rm (avg)}$, as a function of the laser-QD detuning. \AM{Due to the spin not being initialized, the only non-zero component of $\vec S^{\rm (avg)}$ is $s_{HV}^{\rm (avg)}$ \cite{Mehdi2024}.} For large \AM{laser-QD} detunings, $s_{HV}^{\rm (avg)}\rightarrow 1$, as the $H$-polarized input light is directly reflected without interacting with the \LL{cavity nor with the QD}. When the input field is in resonance \AM{with the QD}, $s_{HV}^{\rm (avg)}$ decreases, as a signature of a polarization rotation \cite{Mehdi2024, Gundin2025}. In particular, there are two frequencies for which $s_{HV}^{\rm (avg)}=0$, corresponding to the same detunings which, in Fig.~\ref{fig:1_principle_experiment}c, led to the prediction that $\braket{\Psi_\uparrow|\Psi_\downarrow}=0$ (neglecting decoherence processes).\ The vertical dashed line denotes the positive, optimal detuning $\Delta\omega_{\rm opt}/2\pi=0.14\rm~GHz$ used in this work. As it is closer to resonance with the cavity mode $H$, this detuning provides a higher count rate, both in the $I_{H\rightarrow H}^{\rm (avg)}$ and $I_{H\rightarrow V}^{\rm (avg)}$ intensities, leading to a faster coincidence rate in the correlation measurements. In the following, we fix the laser frequency so that $\Delta\omega=\Delta\omega_{\rm opt}$.\\
	
	\begin{figure}[t!]
		\centering
		\includegraphics[scale=1]{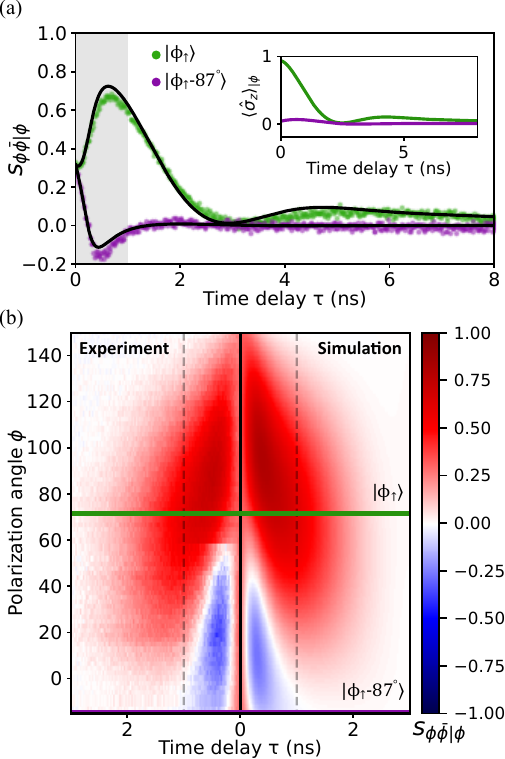}
		\caption{(a) Conditional Stokes parameter $s_{\phi\bar\phi|\phi}(\tau)$ as a function of the delay $\tau$ after a photon detection in the measurement bases $\ket{\phi_\uparrow}$ (green) and $\ket{\phi_\uparrow-87\degree}$ (purple). Experimental data (scatter plot) are compared to numerical simulations of the QD-cavity system. The inset corresponds to the simulated conditional spin component $\braket{\hat\sigma_z}_{|\phi}(\tau)$ as a function of the delay for each basis. (b) Time-dependent conditional Stokes parameter $s_{\phi\bar\phi|\phi}(\tau)$ (colorbar) as a function of the measurement polarization angle $\phi$ (vertical axis) and the time delay (horizontal axis). Experimental data (Left) are compared with simulations (Right). The horizontal lines correspond to the $\ket{\phi_\uparrow}$ (green) and $\ket{\phi_\uparrow -87\degree}$ (purple) measurement bases shown in panel (a). The \AM{vertical dashed lines at $\tau=1\rm~ns$ delimit the timescale governed by the transient regime}. Spin projection effects are read for $\tau\gtrsim 1\rm~ns$. }
		\centering 
		\label{fig:3_2D_Map}
	\end{figure}
	
	\noindent\textbf{Efficient spin projection}
	
	To measure the conditional Stokes parameter $s_{\phi\bar\phi|\phi}(\tau)$ we use Polarimeter I (Fig.~\ref{fig:2_exp_setup_characterization}a). A first photon is measured in polarization $\phi$ and time-correlated with another detector, measuring a second photon after a given delay $\tau$ in the same polarization, extracting the conditional probability $P_{\phi|\phi}(\tau)$, or in the orthogonal one, extracting $P_{\bar\phi|\phi}(\tau)$. The results are shown in Fig.~\ref{fig:3_2D_Map}a, where $s_{\phi\bar\phi|\phi}(\tau)$ is plotted as a function of the delay, after the first photon is detected in polarization $\ket\phi$, with $\phi=\phi_\uparrow=72\degree$ (green) or in polarization $\phi=\phi_\uparrow-87\degree=-15\degree$ (purple). The data are compared to numerical simulations (solid black lines) \AM{including optical pure dephasing and electron hyperfine interaction, while also taking into account charge co-tunneling} \LL{(see Methods and ~\cite{Suppl})}. Both hyperfine interaction and charge co-tunneling play a role in the spin evolution back to equilibrium and the corresponding evolution of the device optical response. \LL{Charge co-tunneling also results in a non-unity charge occupation probability $P_{\rm charge}=0.96\pm0.02$ in the stationary regime.}
	
	In both cases, the value of $s_{\phi\bar\phi|\phi}$ for $\tau=0$ is the same. Then, $s_{\phi\bar\phi|\phi}$ evolves up to $\tau\sim1\rm~ns$ in a regime associated to the time required for the reflection amplitudes to reach their stationary  values under a CW drive. Indeed, a fully-coherent optical response can only be obtained in presence of a monochromatic CW excitation field, in a regime where the dipoles associated to the  $\ket\uparrow-\ket{\Uparrow\downarrow\uparrow}$ and $\ket\downarrow-\ket{\Downarrow\uparrow\downarrow}$ optical transitions are oscillating at the laser frequency~$\omega$. Building such a dipole, after an abrupt modification of the system's density matrix following a photon detection event, takes a time governed by the radiative lifetime $T_1^{\rm (rad)}=200\pm20\rm~ps$, estimated \LL{from} the QD and cavity parameters \cite{Suppl}. Afterwards, we reach $s_{\phi\bar\phi|\phi}\sim0.7$ for $\phi=\phi_\uparrow$, as a signature of efficient spin projection. The conditional Stokes parameter $s_{\phi\bar\phi|\phi}(\tau)$ then decays to 0 before a partial recovery, consistent with a spin relaxation phenomenon governed by the interaction with the Overhauser field \cite{Merkulov2002}, with a \AM{characteristic} timescale $T_1^{\rm (spin)}=1.9\pm0.2\rm~ns$ \cite{Suppl}. Finally, $s_{\phi\bar\phi|\phi}$ fully decays at $\tau\sim10\rm~ns$, associated to the charge co-tunneling regime in which the device operates \AM{with escape time $\tau_{\rm esc}=4.1\pm0.2\rm~ns$} \cite{Medeiros2026}. These features are lost for $\phi=\phi_\uparrow-87\degree$, where $s_{\phi\bar\phi|\phi}\sim0$ after the radiative regime, demonstrating the measurement-dependent response.
	
	The numerically-simulated, conditional spin evolution is shown in the inset of Fig.~\ref{fig:3_2D_Map}a for both polarizations, through $\braket{\hat\sigma_z}_{|\phi}(\tau)$. When the measurement basis is set to maximize the measurement-induced back-action, i.e. \mbox{$\phi=\phi_\uparrow$}, the conditional spin polarization $\braket{\hat\sigma_z}_{|\phi_\uparrow}(0)$ is drastically modified compared to the average spin polarization $\braket{\hat\sigma_z}=0$. \AM{The conditional spin polarization is maximal \LL{immediately} after the detection of the first reflected photon}, in contrast to $s_{\phi\bar\phi|\phi}$, whose evolution rate at short delays is governed by the radiative lifetime $T_1^{\rm (rad)}$. Based on numerical simulations \cite{Suppl}, we estimate a spin population imbalance at zero delay of $\braket{\hat\sigma_z}_{|\phi_\uparrow}(0)=0.94\pm0.02$, with conditional probabilities $P_{\uparrow|\phi_\uparrow}=95\pm2\%$ and $P_{\downarrow|\phi_\uparrow}=3\pm2\%$, leading to a total conditional occupation probability $P_{\rm charge|\phi_\uparrow}=98\pm2\%$. The discrepancy between $P_{\rm charge|\phi_\uparrow}$ and the unconditional charge probability $P_{\rm charge}=96\pm2\%$ is expected, since the detection of a $\phi_\uparrow$-polarized photon is more likely if the QD is charged, increasing $P_{\rm charge|\phi_\uparrow}$. Still, $P_{\rm charge|\phi_\uparrow}$ remains slightly lower than unity due to the possibility that an $H$-polarized photon directly reflected by an uncharged device is measured in the $\ket{\phi_\uparrow}$ state. The residual probability to be in the $\ket\downarrow$ state, i.e. $P_{\downarrow|\phi_\uparrow}$, is attributed to incoherent processes such as pure dephasing and electron hyperfine interaction, both leading to a \LL{slight} broadening of the optical transition and an averaging of the $\vec S_\uparrow$ and $\vec S_\downarrow$ states, which then do not fulfill the optimal condition $\vec S_\uparrow=-\vec S_\downarrow$, with $\|\vec S_\uparrow\|=\|\vec S_\downarrow\|=1$.
	
	In contrast, for $\phi=\phi_\uparrow-87\degree$ the spin remains almost mixed due to an inefficient spin population imbalance, as now $\phi\approx\phi_\uparrow-90\degree$. This leads to $\braket{\hat\sigma_z}_{|\phi}(0)\sim0$, corresponding to $P_{\uparrow|\phi}=51\pm2\%$ and $P_{\downarrow|\phi}=47\pm2\%$, i.e. an almost complete lack of knowledge on the spin state following photon detection. The slight imbalance is mainly attributed to not measuring in $\phi_\uparrow-90\degree$, due to experimental limitations in the calibration of the measurement bases.
	
	Fig.~\ref{fig:3_2D_Map}b shows the time-dependence of the conditional Stokes parameter $s_{\phi\bar\phi|\phi}(\tau)$ as a function of the measurement basis angle $\phi$. For each basis (vertical axis) we measure the time evolution (horizontal axis) of $s_{\phi\bar\phi|\phi}(\tau)$ (color scale). The experimental data (Left) is compared to numerical simulations (Right). The horizontal lines represent the cross-sections shown in Fig.~\ref{fig:3_2D_Map}a for the measurement polarization states $\ket{\phi_\uparrow}$ (green) and $\ket{\phi_\uparrow - 87\degree}$ (purple). The \AM{vertical dashed lines at $\tau=1\rm~ns$ refer to the timescale below which we evidence the influence of the transient regime}, governed by the radiative lifetime $T_1^{\rm (rad)}$, corresponding to the time it takes to build coherent superpositions of the ground and excited states after \AM{a photon detection has projected the spin}, for both optical transitions in Fig.~\ref{fig:1_principle_experiment}a. Spin projection features are read for $\tau \gtrsim 1\rm~ns$, for each polarization angle $\phi$. When the measurement basis is set to $\phi=\phi_\uparrow-87\degree$, no spin projection features are observed ($s_{\phi\bar\phi|\phi}\sim0$ at $\tau>1\rm~ns$ as seen in Fig.~\ref{fig:3_2D_Map}a). When the basis is gradually changed, $s_{\phi\bar\phi|\phi}$ increases in magnitude and therefore the spin dynamics effects \AM{can be resolved}. For $\phi=\phi_\uparrow$, the spin is maximally projected, with its relaxation dynamics mapped onto the evolution of $s_{\phi\bar\phi|\phi}$.\\
	
	\noindent\textbf{Spin-to-polarization mapping}
	
	While we have shown that the measurement of $s_{\phi\bar\phi|\phi}(\tau)$ informs us on the measurement-induced projection of the QD-hosted spin, we aim to determine the $\vec S_\uparrow$ and $\vec S_\downarrow$ Stokes vectors in order to efficiently operate the device. These Stokes vectors, however, are not directly accessible in absence of spin pumping. Instead, we leverage our spin heralding approach and measure the polarization state $\vec S_{|\phi}(\tau)$ of a second reflected photon, \LL{conditioned to} the measurement of a first one in the polarization state $\ket\phi$. \LL{In a simplified model}, the conditional Stokes vector $\vec S_{|\phi}(\tau)$ \LL{can be} interpreted as:
	\begin{equation}
		\begin{split}
			\vec S_{|\phi}(\tau)&=P_{\uparrow|\phi}(\tau)\vec S_\uparrow+P_{\downarrow|\phi}(\tau)\vec S_\downarrow,
		\end{split}
		\label{eq:conditional_stokes_vector_simpl}
	\end{equation}
	with $P_{\uparrow|\phi}(\tau)$ (resp. $P_{\downarrow|\phi}(\tau)$) the conditional probability of finding the spin in the $\ket\uparrow$ (resp. $\ket\downarrow$) state after detecting a photon in polarization $\phi$. \AM{For an ideal interface and in the absence of spin relaxation}, if the spin is heralded in the $\ket\uparrow$ state with $\phi=\phi_\uparrow$ ($P_{\uparrow|\phi_\uparrow}=1$), this leads to a direct measurement of $\vec S_\uparrow$, with $\vec S_{|\phi_\uparrow}=\vec S_\uparrow$. The same applies for the $\ket\downarrow$ state with a first photon measured in the state $\ket{\bar\phi}$ with $\bar\phi=\phi_\uparrow+180\degree$, ideally leading to $P_{\downarrow|\bar\phi}=1$ and thus to $\vec S_{|\phi_\downarrow}=\vec S_\downarrow$.
	
	\begin{figure*}[t!]
		\includegraphics[scale = 1]{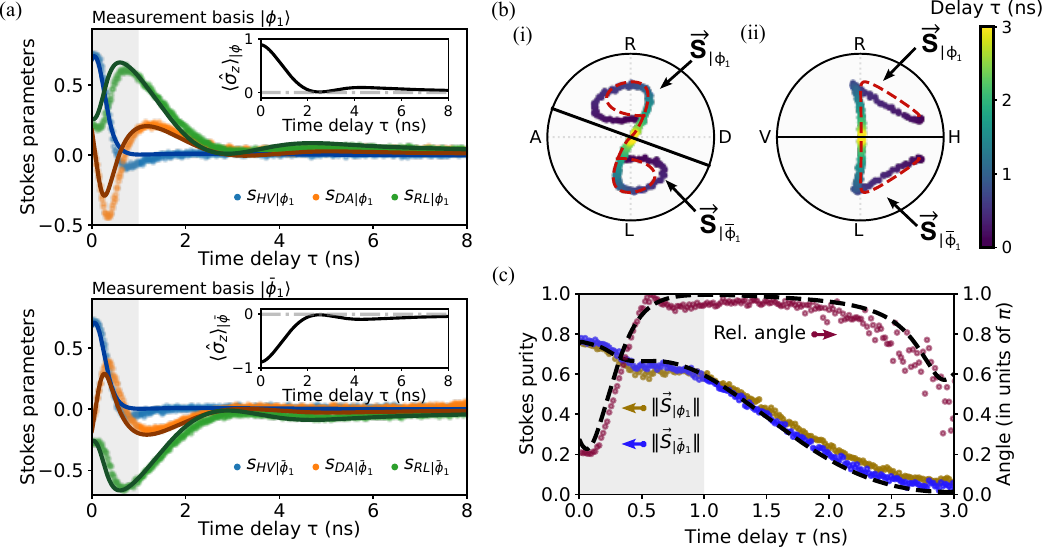}
		\caption{(a) Time-dependent conditional Stokes parameters $s_{\rm HV|\phi_1}$ (blue), $s_{\rm DA|\phi_1}$ (orange) and $s_{\rm RL|\phi_1}$ (green) measured after a first detected photon in the measurement bases $\ket{\phi_1}$ (Top) and $\ket{\bar{\phi_1}}$ (Bottom), with $\phi_1=\phi_\uparrow-16\degree$. The experimental data (scatter plot) are compared to numerical simulations (solid lines). The inset of each panel shows the predicted evolution of the conditional spin polarization $\braket{\hat{\sigma}_z}_{|\phi}$.
			(b) Poincaré sphere representation of the conditional Stokes vectors $\vec S_{|\phi_1}$ and $\vec S_{|\bar{\phi_1}}$ for the cross-sections \mbox{(i) DARL} and \mbox{(ii) HVRL} as a function of time. Experimental data (scatter plots) are compared to numerical simulations (dashed red lines). (c) Left axis: Polarization state purities $\|\vec S_{|\phi_1}\|$ and $\|\vec S_{|\bar{\phi_1}}\|$ for measurement basis $\ket{\phi_1}/\ket{\bar{\phi_1}}$ as a function of the delay. Right axis: Relative angle between the Stokes vector $\vec{S}_{|\phi}$ and $\vec{S}_{\bar{\phi}}$. Each experimental plot is compared to numerical simulations.}
		\centering
		\label{fig:4_conditional_tomography}
	\end{figure*}
	
	In the following, we display conditional polarization tomography measurements taken with a first photon measured in the polarization states $\ket{\phi_1}$ and $\ket{\bar{\phi_1}}$, with $\phi_1=56\degree$ and $\bar{\phi_1}=236\degree$. This measurement basis was chosen before numerical simulations were made: the angles $\phi_1$ and $\bar{\phi_1}$ thus slightly differ from the optimal angles $\phi_\uparrow$ and $\phi_\downarrow$, yet this only diminishes by a few percent the efficiency of the spin projection. The time evolution of the polarization states $\vec S_{|\phi_1}(\tau)$ and $\vec S_{|\bar{\phi_1}}(\tau)$ are shown in Fig.~\ref{fig:4_conditional_tomography}a through the conditional Stokes parameters $s_{HV|\phi}$, $s_{DA|\phi}$ and $s_{RL|\phi}$ for $\phi=\phi_1$ (Top panel) and $\phi=\bar{\phi_1}$ (Bottom panel). To measure these quantities, a first photon is measured in the $\phi_1$ or $\bar{\phi_1}$ polarization with Polarimeter~I to herald the spin state. Then, Polarimeter II performs the conditional tomography of a second reflected photon (see Fig.~\ref{fig:2_exp_setup_characterization}a). At short delays ($\tau\lesssim1\rm~ns$), we first observe that $s_{HV|\phi_1}$ and $s_{HV|\bar{\phi_1}}$ strongly differ from $s_{HV}^{\rm (avg)}=0$, yet rapidly decay back to this steady-state value in a timescale governed by the radiative lifetime $T_1^{\rm (rad)}$. These dynamics are spin independent, since $s_{HV|\phi_1}\approx s_{HV|\bar{\phi_1}}$ all along the evolution, and quickly decay to 0 when  $\tau\sim1\rm~ns$. Alongside the evolution of $s_{DA|\phi}$ and $s_{RL|\phi}$ \LL{at short delays}, this provides a polarization-resolved measurement of the transient optical response of the device, associated to the time it takes to establish an electric dipole for each optical transition, after the initial back-action event.
	
	For $\tau\gtrsim 1\rm~ns$, the conditional Stokes parameters $s_{DA|\phi_1}$ and $s_{RL|\phi_1}$ are almost opposed to $s_{DA|\bar{\phi_1}}$ and $s_{RL|\bar{\phi_1}}$, respectively. This is a signature of a spin-dependent polarization rotation, since detecting a photon in polarization $\phi_1$ ($\bar{\phi_1}$) heralds the spin in the $\ket\uparrow$ ($\ket\downarrow$) state  with $P_{\uparrow|\phi_1}=P_{\downarrow|\bar{\phi_1}}=93\pm2\%$ (\LL{numerically} extracted from the same set of device parameters previously discussed\LL{, see Methods and \cite{Suppl}}). The simulated conditional spin state $\braket{\hat\sigma_z}_{|\phi}(\tau)$ is shown in the inset of each plot for $\phi=\phi_1$ and $\phi=\bar{\phi_1}$, with $|\braket{\hat\sigma_z}_{|\phi}(0)|=0.90\pm0.02$. This leads to non-zero conditional Stokes parameters $s_{DA|\phi_1}$ and $s_{RL|\phi_1}$ (resp. $s_{DA|\bar{\phi_1}}$ and $s_{RL|\bar{\phi_1}}$), whose relative amplitudes inform us on the orientation of $\vec S_\uparrow$ (resp. $\vec S_\downarrow$). At larger delays, all these conditional Stokes parameters both decay towards zero, i.e. towards $s_{DA}^{\rm (avg)}$ and $s_{RL}^{\rm (avg)}$, following spin relaxation.
	
	The time-evolution of the $\vec S_{|\phi_1}(\tau)$ and $\vec S_{|\bar{\phi_1}}(\tau)$ is displayed in the (i) $DARL$ and (ii) $HVRL$ cross-sections of the Poincaré sphere in Fig.~\ref{fig:4_conditional_tomography}b, as a function of the delay. The experimental data (scatter plot) are compared to numerical simulations (red dashed line), while the colorscale represents the time after the first photon detection in polarization $\phi_1$ or $\bar{\phi_1}$ (a 3D view of these plots is available in \cite{Suppl}). Each cross-section is separated in two hemispheres to separate the dynamics of the $\vec S_{|\phi_1}$ (Top hemisphere) and the $\vec S_{|\bar{\phi_1}}$ (Bottom hemisphere) states. In both cross-sections, for $\tau\gtrsim1\rm~ns$ i.e. after the transient regime (whose evolution is well captured by our model), both vectors decay towards the center of the sphere. In (i),  the $\vec S_{|\phi_1}$ and $\vec S_{\bar{\phi_1}}$ vectors are seen evolving along the same axis, tilted from the $DA$ axis by an angle of $\phi_\uparrow=72\degree$, with opposite orientations. In (ii), both vectors relax to $s_{HV|\phi_1}=0$. Taken together, these cross-sections indicate that, after the transient regime, $\vec S_{|\phi_1}$ and $\vec S_{|\bar{\phi_1}}$ become perpendicular to the $HV$ axis and opposed to one another. To interpret these observations, one can use the simplified model captured by Eq.~(\ref{eq:conditional_stokes_vector_simpl}). Since the conditional spin projection probabilities are reversed for the two measurement bases, i.e. $P_{\uparrow|\phi_1}\approx P_{\downarrow|\bar{\phi_1}}$, we retrieve $\vec S_{|\phi_1}=-\vec S_{|\bar{\phi_1}}$ only if $\vec S_\uparrow=-\vec S_\downarrow$. In such a case, we expect $\vec S_{|\phi_1}$ and $\vec S_{|\bar{\phi_1}}$ to follow symmetrical evolutions along the axes determined by $\vec S_\uparrow/\vec S_\downarrow$, and decay towards the mixed state as the conditional spin probabilities converge to 1/2. This simplified picture can only be applied, however, after the transient regime within which the spin-dependent vectors $\vec S_\uparrow$ and $\vec S_\downarrow$ are not yet stabilized. For an electron spin, $T_1^{\rm (spin)}$ is comparable to $T_1^{\rm (rad)}$, and so the conditional spin probabilities are already decreased at $\tau\sim1\rm~ns$.
	
	A quantitative study of this dynamics is done in Fig.~\ref{fig:4_conditional_tomography}c, where the relative angle $\alpha$ between the $\vec S_{|\phi_1}$ and $\vec S_{|\bar{\phi_1}}$ Stokes vectors is shown together with the vector purities $\|\vec S_{|\phi_1}\|$ and $\|\vec S_{|\bar{\phi_1}}\|$, as a function of the delay. We find $\alpha/\pi=0.95\pm0.02$ ($\alpha=171\degree\pm4\degree$) for the range of delays between $\tau=1\rm~ns$ and $\tau=2\rm~ns$, where the polarization purity decreases from $60\%$ to $20\%$. Following Eq.~(\ref{eq:conditional_stokes_vector_simpl}), the fact that the relative angle $\alpha$ is close to $\pi$ (i.e. $\vec S_{|\phi_1}\approx-\vec S_{|\bar{\phi_1}}$) implies that $\vec S_\uparrow\approx-\vec S_\downarrow$, as required for an optimal spin-photon interface. The decrease of $\|\vec S_{|\phi_1}\|$ and $\|\vec S_{|\bar{\phi_1}}\|$ is then understood as a decay of the conditional spin probabilities $P_{\uparrow/\downarrow|\phi}$ towards the mixed state, leading $\vec S_{|\phi_1}$ and $\vec S_{|\bar{\phi_1}}$ to decay towards the fully mixed state (at the center of the sphere) if $\vec S_\uparrow=-\vec S_\downarrow$. We attribute the discrepancy from the perfect antiparallel case to experimental noise, like pure dephasing and hyperfine interaction.\\
	
	\noindent \textbf{\Large{Discussion}}
	
	In conclusion, we leverage the spin-induced Kerr rotations present in a charged QD-micropillar cavity device to engineer a spin-to-polarization mapping, whereby two orthogonal spin eigenstates lead to almost orthogonal polarization states, associated to almost opposite Stokes vectors, for the reflected photons. We first leverage this mapping to efficiently project the spin of a trapped electron, upon the detection of a single reflected photon, with an estimated probability to be in the $\ket\uparrow$ state of $P_{\uparrow|\phi}=95\pm2\%$. The spin dynamics is then transferred to the polarization state of a second reflected photon, analyzed through time-resolved conditional polarization tomography measurements.
	
	The measured data evidence the important role played by the optical transient regime, with the timescale \mbox{$T_1^{\rm (rad)}=200\pm20\rm~ps$}, during which the device coherent optical response is built, allowing the spin-dependent Stokes vectors $\vec S_\uparrow$ and $\vec S_\downarrow$ to reach their stationary values. As we do not directly measure $\vec S_\uparrow$ and $\vec S_\downarrow$, but instead conditional Stokes vectors $\vec S_{|\phi}$ and $\vec S_{|\bar\phi}$ heralded by a first photon detection, our results are also strongly impacted by spin relaxation, mainly governed by the electron hyperfine interaction, at a timescale $T_1^{\rm (spin)}=1.9\pm0.2\rm~ns$. This relaxation reduces the conditional probabilities $P_{\uparrow|\phi}$ and $P_{\downarrow|\phi}$, limiting the purity of the reflected conditional Stokes vectors $\|\vec S_{|\phi}\|$ and $\|\vec S_{|\bar\phi}\|$. The timescale competition between the radiative lifetime and the spin relaxation time is therefore the current main limitation of our approach, which can be overcome by the use of different device structures improving on the radiative lifetime \cite{Margaria2025} and/or the spin properties (such as by using a hole spin \cite{Coste2023, Hogg2025}). In these conditions, low-noise QD-cavity devices can be operated as optimal spin-photon interfaces, leading to pure spin-dependent polarization states satisfying $\braket{\Psi_\uparrow|\Psi_\downarrow}=0$.
	
	This work paves the way towards deterministic logical gates, such as a C-NOT gate \cite{Bonato2010, Wei2013}, where the polarization of an incoming photon is flipped in a spin-dependent way. More generally, the realization of a process map $\ket\uparrow\rightarrow\ket\uparrow\ket{\Psi_\uparrow}$ and $\ket\downarrow\rightarrow\ket\downarrow\ket{\Psi_\downarrow}$, with $\braket{\Psi_\uparrow|\Psi_\downarrow}=0$ is a core requirement to use coherent spin-photon interfaces as entanglers of incoming photons \cite{Hu2008_Entangler} to potentially realize multi-dimensional graph states using a single spin \cite{Zhan2020}.\\

	\noindent \textbf{\Large{Methods}}
	
	\noindent \textbf{Sample design}
	
	The QD-cavity device was grown by molecular beam epitaxy. An annealed InGaAs QD is embedded in a $\lambda$-GaAs cavity, formed by two distributed Bragg reflectors consisting of alternating layers of GaAs and Al$_{0.9}$Ga$_{0.1}$As, with 20 (30) pairs for the top (bottom) mirror. The bottom mirror presents a gradual doping (Si-doped) from $2\times10^{18}$~cm$^{-3}$ down to $1\times10^{18}$~cm$^{-3}$, to electrically contact the structure. This level of doping is maintained in the first half of the cavity region and is stopped 25~nm before the QD layer, creating a tunnel barrier between the quantum dot and the Fermi sea. The top mirror is increasingly C-doped, from zero to $2\times10^{19}$~cm$^{-3}$ at the surface. Four ridges connects the micropillar to a large circular frame, attached to a gold-plated mesa enabling the electrical control. For full details on the device fabrication, see \cite{Somaschi2016,DeSantis2017}.\\
	
	\noindent \textbf{Device characterization}
	
	The QD-cavity device is placed in a liquid Helium cryostat at 4K (attoDry1000) and optically addressed with a tunable CW laser in the low power regime ($P_{\rm in}=10\rm~pW$). Light is focused and collected using a cold aspheric lens (4mm focal length). A single mode fiber prepares the spatial profile of the input beam, and its diameter adjusted with a zoom lens to focus it into the device \cite{Hilaire2018}. By means of polarization-resolved scattering measurements \cite{Mehdi2024}, we characterize the two cavity modes $H$ and $V$: the modes are separated by $\omega_H-\omega_V = 2\pi\times(32.2\pm0.5)\rm~GHz$, with linewidths $\kappa_H/2\pi=44\pm2\rm~GHz$ and \mbox{$\kappa_V/2\pi=47\pm1\rm~GHz$}. The extraction efficiency through the top mirror is $\eta_\mathrm{top}=0.63\pm0.01$. 
	
	To address the QD, \AM{emitting at 925nm,} we perform the same type of measurements with an applied bias voltage of \mbox{$V=-1.2\rm~V$}, and the data are compared to numerical simulations to extract the relevant parameters. The QD is slightly blue-detuned from the center cavity frequency by $6.3\pm0.7\rm~GHz$, with light-matter coupling strength $g/2\pi=3.1\pm0.2\rm~GHz$, spontaneous decay rate outside the cavity modes $\gamma_{\rm sp}/2\pi=160\pm15\rm~MHz$ and pure dephasing rate $\gamma^*/2\pi=24\pm10\rm~MHz$.
	
	The dynamics measured with the two-photon correlations are compared with numerical simulations, where we extract the coupling constant $\gamma_e/2\pi=120\pm12\rm~MHz$ for electron hyperfine interaction (we neglect hole hyperfine interaction) and co-tunneling parameters with capture and escape time $\tau_{\rm cap}=0.2\pm0.1\rm~ns$ and $\tau_{\rm esc}=4.1\pm0.2\rm~ns$ respectively, leading to an average charge occupation probability $P_{\rm charge}=0.96\pm 0.02$.
	
	\AM{For the simplified model discussed in Fig.~\ref{fig:1_principle_experiment}c, we use the cavity parameters $\omega_{H(V)}$, $\kappa_{H(V)}$ and $\eta_{\rm top}$, alongside with the QD resonance frequency $\omega_{\rm QD}$, $g$ and $\gamma_{\rm sp}$. For the rest of the figures, where numerical simulations are performed, we use the aforementioned parameters including optical pure dephasing with rate $\gamma^*$, as well as spin relaxation with rate $\gamma_{\rm e}$ and charge dynamics with $\tau_{\rm cap}$, $\tau_{\rm esc}$ and $P_{\rm charge}$.}\\

	\noindent \textbf{ACKNOWLEDGEMENTS} \\
	This work was partially supported by the Paris Ile-de-France Région in the framework of DIM SIRTEQ, the Research and Innovation Programme QUDOT-TECH under the Marie Sklodowska-Curie grant agreement 861097, Horizon CL4 program under the
	grant agreement 101135288 for EPIQUE project, the European Union’s Horizon 2020 FET OPEN project QLUSTER (Grant ID 862035), the French National Research Agency (ANR) through projects \AM{ANR-22-PETQ0011} and ANR-22-PETQ0013, a public grant as part of the ”Investissements d’Avenir” programme (Labex NanoSaclay, reference: ANR10LABX0035) and a government grant as part of France 2030 (QuanTEdu-France), under reference ANR-22-CMAS-0001. \AM{This work was conducted within the research program of the QDLight joint laboratory (C2N/Quandela).} This work was done within the C2N micro nanotechnologies platforms and partly supported by the RENATECH network and the General Council of Essonne.\\
	
	\noindent \textbf{AUTHOR CONTRIBUTIONS} \\
	A.M. and V.V. performed the experiments. \LL{A.M., V.V., E.M. and D.A.F. built and calibrated the optical setup}. A.M. analyzed the data. A.L., I.S., N.S. \LL{and P.Stp. prepared} the device based on a design by P.S. Numerical simulations were developed by E.M., C.M., M.G. and A.M. The analytical model was developed and tested by A.M. and E.R. A.M. prepared all figures and wrote the manuscript. P.St., O.K., P.S. and D.A.F. helped with the discussion of scientific methods and data analysis. L.L. designed the project and supervised the experiments, data analysis and manuscript writing. All authors discussed the results and helped with the manuscript preparation.
	
	\noindent \textbf{COMPETING INTERESTS} \\
	N.S. is a co-founder and P.S. is a scientific advisor and co-founder of the company Quandela. The other authors declare no competing interests.\\

\end{document}